\documentclass[a4paper,12pt]{article}
\usepackage[utf8]{inputenc}
\usepackage{amsmath}
\usepackage{amsfonts}
\usepackage{amssymb}
\usepackage{amsthm}
\usepackage{caption}
\usepackage{multirow}
\usepackage{fontenc}
\usepackage{graphicx}
\usepackage{color}
\usepackage{hyperref}
\usepackage[symbol]{footmisc}
\usepackage[total={6in, 9.5in}]{geometry}

\date{}

\def\title#1{\begin{center} {\LARGE #1 \vspace{0.5cm}} \end{center}}
\def\author#1{\begin{center}{ \large #1} \end{center}}
\def\affil#1{\begin{center}{ \it #1} \end{center}}
\def\email#1{\begin{center}{\normalsize #1} \end{center}}
\def\preprint#1{\rightline{\begin{tabular}{l} #1 \end{tabular}}}

\begin{document}
\preprint{ABCD-ABCD} 

\title{Neutrino physics in slowly rotating black hole spacetime and nonlinear electrodynamics} 

\author{G. Lambiase$^{1\,\ast}$\footnote[5]{Speaker}\,\,, H. Mosquera Cuesta$^{2\,\dagger}$\,\,, and J.P. Pereira$^{3\,\ddagger}$} 

\affil{$^1$Dipartimento di Fisica "E.R. Caianiello", Universit\'a
di Salerno, 84081 Baronissi (Sa), Italy. \\ $^2$INFN, Sezione di Napoli, Gruppo Collegato,  Italy  \\
$^2$ Instituto Federal de Educa\c c\~ao, Ci\^encia e Tecnologia
do  Cear\'a, Avenida Treze de Maio, 2081, Benfica, Fortaleza/CE, CEP 60040-531, Brazil \\
$^3$Nicolaus Copernicus Astronomical Center, Polish Academy of Sciences, Bartycka 18, 00-716, Warsaw, Poland}

\email{$^{\ast}$lambiase@sa.infn.it,\, $^{\dagger}$herman.paye@gmail.com,\, $^{\ddagger}$jpereira@camk.edu.pl}
\vspace{1cm}
%
%
\begin{abstract}
Huge electromagnetic fields are known to be present during the late stages of the dynamics of supernovae.
Thus, when dealing with electrodynamics in this context, the possibility may arise to probe nonlinear theories.
The Einstein field equations minimally
coupled to an arbitrary nonlinear Lagrangian of electrodynamics are solved in the regime of slow rotation, i.e. $a \ll M$ (black hole’s mass), up to first order in $a/M$. We use Born-Infeld Lagrangian to compare the obtained results with Maxwellian counterpart.
We focus on the astrophysics of neutrino flavor oscillations
($\nu_e \to \nu_{\mu, \tau}$) and spin-flip ($\nu_L \to \nu_R$), as well as on the computation of that the electron fraction $Y_e$,
hence the r-processes, which may significantly differ with respect to the standard electrodynamics.
\end{abstract}

\vspace*{2.4cm}
\begin{center}
{\LARGE{Presented at}}\\
\vspace{1cm}
\Large{NuPhys2019: Prospects in Neutrino Physics\\
Cavendish Conference Centre, London, 16--18 December 2019}
\end{center}

\clearpage


\section{Introduction}

Astrophysical environments provide a laboratory for studying the evolution of neutrinos.
In these contexts, in fact, neutrino species could find themselves propagating in the spacetime (the gravitational field)
around a (generally rotating and nonlinearly charged) black
hole. In particular, in the case of a collapsing proto-neutron star (P-NS) or a (just-formed) charged black
hole, the spacetime is characterized by a rotation parameter as well as electric and magnetic fields, which play important dynamical roles.
Both fields are therefore crucial for the physics of neutrino interactions in
the very deep interior of a supernova, especially if the electromagnetic field is
overcritical, i.e. one which surpasses the field in which the classical vacuum
breaks down (Schwinger's limit $4.4 \times 10^{13}$ G \cite{scwinger}).

In this paper we discuss some aspects of neutrino physics in spacetimes where nonlinear electrodynamics are taken into account.
We refer to a representative of them as a slowly rotating nonlinear charged black hole (SRNLCBH), and we assume that the nonlinear electrodynamics is described by the Born-Infeld theory (for applications, see \cite{applBI}).
In particular, we apply the generic results of neutrino oscillations and spin precession for the
Born-Infeld theory, in order to explore their differences with respect to the Maxwell Lagrangian.
We also make use of the effect of frame dragging in axially symmetric spacetimes to contrast the aforementioned theories.
Finally, we discuss the relevance of the charge of a black hole (nonlinearity of the
electromagnetism) for r-processes. Simple
estimates are given in order to assess relevant
scales for some physical processes in the astrophysical context.

\section{Nonlinear electrodynamics}

Nonlinear electrodynamics (NLED) is a theory to describe electromagnetic interactions in a relativistically invariant set up.
Several approaches have been envisioned \cite{H-E,born,plebanski}. We shall consider the Born-Infeld model, whose Lagrangian density is given by
\[
L^{B-I} = b^2\left[1-\sqrt{1+\frac{F}{2b^2}-\frac{G}{16b^4}} \right]\,,
\]
where $b$ is the parameter of the model, and
  \[
 F\doteq F^{\mu\nu}F_{\mu\nu}\,, \quad  G\doteq {}^* F^{\mu\nu}F_{\mu\nu}, \quad {}^*F^{\mu\nu} \doteq \frac{1}{2\sqrt{-\det|g_{\mu\nu}|}} \epsilon^{\mu\nu\alpha\beta} F_{\alpha\beta}
\]
The Einstein field equations are given by
\begin{equation}\nonumber
{ G_{\mu\nu}= 8\pi (T_{\mu\nu}^{mat}+T^{NLED}_{\mu\nu})}, \quad T^{NLED}_{\mu\nu}= 4L_FF_{\mu\alpha} F_{\nu\beta}g^{\alpha\beta}-(L-GL_G)g_{\mu\nu}
\end{equation}
where $T^{NLED}_{\mu\nu}$ ($T_{\mu\nu}^{mat}$) is the electromagnetic energy-momentum tensor of the nonlinear electrodynamics (matter content). In our analysis, we take $T_{\mu\nu}^{mat}$ associated with the energy-momentum tensor of a black hole spacetime. The generalized Maxwell field equations are
 \[
  \frac{\partial}{\partial x_{\mu}}[\sqrt{-g}(L_{F}F^{\mu \nu} + L_G {}^* F^{\mu\nu})] =0
  \]
 \[
\frac{\partial}{\partial x_{\mu}}(\sqrt{-g} {}^* F^{\mu\nu})=0\,, \quad  L_F=\frac{\partial L}{\partial F}\,,\,\,\,  L_G=\frac{\partial L}{\partial G}
\]
We assume that the background geometry is described by Kerr geometry ($a=J/M$ is the rotation parameter)
\[
ds^2=g_{00}dt^2+g_{ij}dx^i dx^j-2 a \sin^2 \theta A(r) dt d\phi 
\]
in which the time-time metric component is \cite{breton}
\begin{equation}\nonumber
g_{00} = 1- 2u + \frac{2}{3u^2}{(bM)}^2\left(1-\sqrt{1+ \frac{\alpha^2u^4}{{(bM)}^2}} \right) +
 \frac{2\alpha^2u}{3}\sqrt{\frac{{bM}}{|\alpha|}} F[z, \frac{1}{\sqrt{2}}];
\end{equation}
here, $F[z, \frac{1}{\sqrt{2}}]=F\left[ \arccos\left({\frac{{bM}- |\alpha|u^2}{{bM} + |\alpha|u^2}}\right),\frac{1}{\sqrt{2}}\right]$ is the elliptic function of first kind, $u=M/r$ and $\alpha=Q/M$. The parameter  $b$ is chosen such that $bM \geqslant 10^{-4}$, which is of interest for astrophysical applications. The function $A(r)$ is determined using the NLED field equations.

\section{Main results}

In this Section we report the main results of neutrino flavor oscillations and spin-flip oscillations in a SRNLCBH \cite{lambiase}.

\vspace{0.3cm}

\centerline{\it Neutrino flavor oscillations}

\vspace{0.15cm}


Neutrino flavor oscillations take place since neutrino flavor eigenstates ($|\nu_{\alpha}\rangle$) and neutrino mass eigenstates ($|\nu_{j}\rangle$) are mixed
\begin{equation}\nonumber
|\nu_{\alpha}\rangle= U_{\alpha j}\exp[-i\Phi_j]|\nu_{j}\rangle,\quad \Phi_j= \int g_{\mu\nu}(x)p_{(j)}^\mu dx^{\nu}
\end{equation}

The transition probability for two-flavored neutrinos is \cite{cardall}
\begin{equation}\nonumber
P(\nu_{\alpha}\rightarrow \nu_{\beta})= \sin^2(2\Theta)\sin^2\left(\frac{\Phi_{jk}}{2}\right), \quad \Phi_{jk}\doteq \Phi_j-\Phi_k
\end{equation}

\begin{equation}\nonumber
\Phi_j = \int_{r_{in}}^{r_{fin}} dr\frac{m_j}{\dot{r}}= m_j^2 \int \frac{dr}{\sqrt{E^2 - g_{00}(r) \left[\frac{l^2}{\sin^2 \theta_0 r^2}- \frac{2E laA(r)}{g_{00}(r)r^2} + m_j^2 \right]}}
\end{equation}
Therefore, the oscillation length is related to the phase difference $L_{osc} \sim \Delta \Phi_{jk}$. For SRNLCBH one gets
\begin{equation}\nonumber
L_{osc}  \doteq \frac{dl_{pr}}{d\Phi_{jk}/(2\pi)} = \frac{2\pi E}{\sqrt{g_{00}} (m^2_j-m^2_k)}\,, \quad dl^2_{pr}=\left(-g_{ij} + \frac{g_{0i}g_{0j}}{g_{00}} \right)dx^idx^j\,.
\end{equation}

\vspace{0.3cm}

\centerline{\it Neutrino spin precession}

\vspace{0.15cm}

The equations governing the spin $S^{\mu}$ coupling of test particles with the gravitational field are \cite{Dvornikov}
\begin{equation}\nonumber
\frac{D S^{\mu}}{d\lambda}=0,\;\; \frac{D u^{\mu}}{d\lambda}=0 ,
\end{equation}
In the rest frame of the particles
\begin{equation}\nonumber
\frac{d{\boldsymbol \xi}}{d\tau}= [\boldsymbol{\xi}\times \boldsymbol{\varpi}],\;\;\quad \boldsymbol{\varpi}\doteq \left(\vec{B} + \frac{\vec{E}\times \vec{u}}{1+u^0}\right)\,.
\end{equation}
\begin{equation}\nonumber
E_i=\varpi_{0i},\;\; B_k=\frac{1}{2}\,\epsilon_{ijk} \varpi_{ij},\quad \varpi^{ab}\doteq \eta^{ac}\eta^{bd}{ \gamma_{cde}}u^e\,, \,\, { \gamma_{abc}\doteq e_{a\mu;\nu}e^{\mu}_be^{\nu}_c}\,.
\end{equation}
The probability of spin-flip (s.f.) for neutrinos in a slowly, rotating and charged spacetime is given by
\begin{equation}\nonumber
{\cal P}_{s.f.}(t)=\sin^2 (|\Delta \boldsymbol{\varpi}|\tau)\,, \quad (\Delta \boldsymbol{\varpi})^2=\frac{\partial_r g_{00}}{2r}\,. 
\end{equation}

\vspace{0.3cm}

\centerline{\it Neutrino oscillations and spin-flip in B-I Lagrangian}

\vspace{0.15cm}

Results of the previous subsection (neutrino flavor and spin-flip oscillations) are reported in Fig. 1. Here the B-I oscillation length and spin-flip frequency are compared with the Maxwell theory.

\begin{figure}[h!]
\centering
 \includegraphics[height=8.cm,width=14.cm]{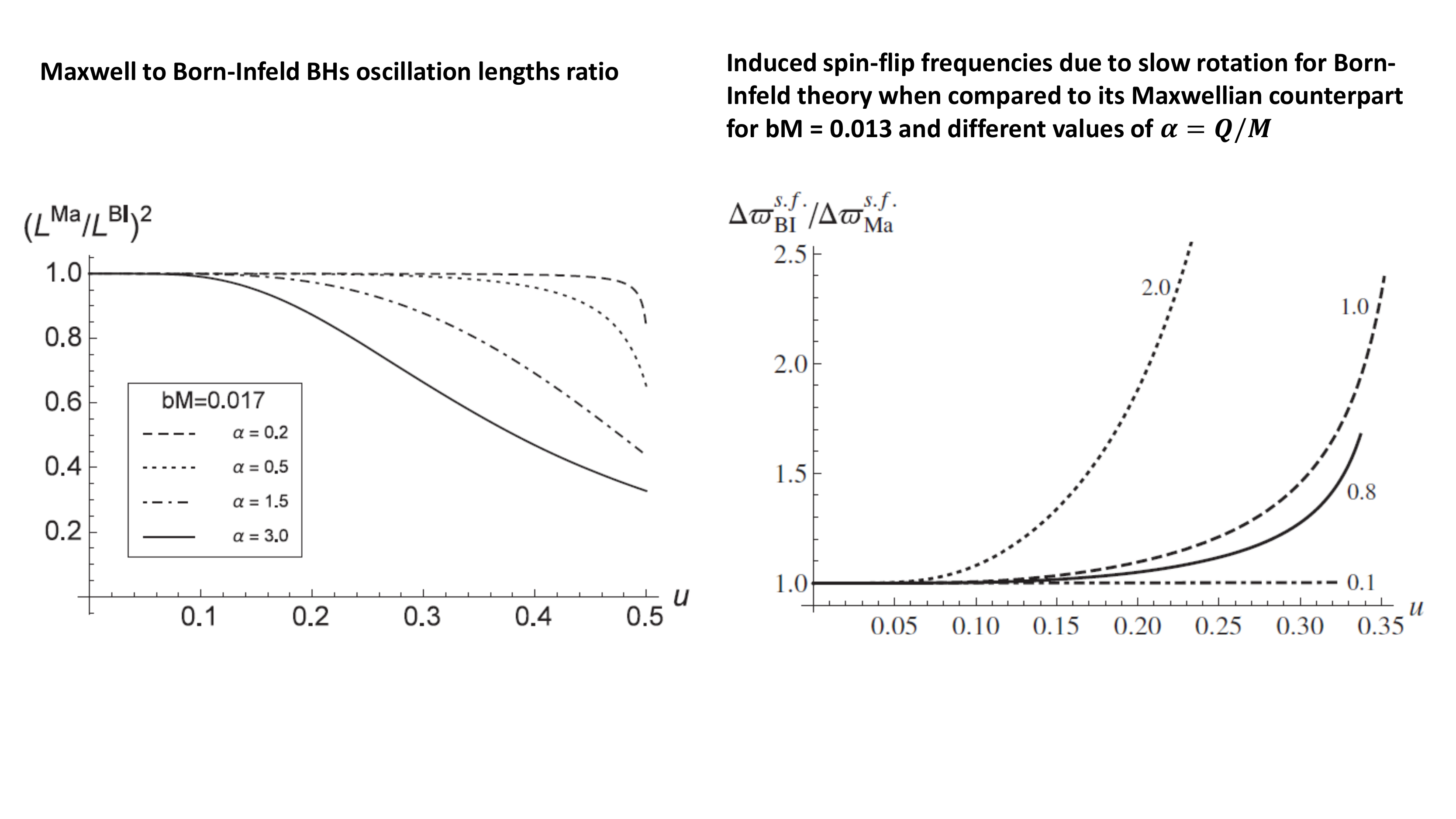}
  \label{Fig1} \vspace{-1.7cm}
  \caption{Neutrino oscillation lengths and spin-flip frequency comparison between the Maxwell theory and Born-Infeld theories.}
 \end{figure}

\subsection{\bf Electron fraction $Y_e$ in NLED-BH}

Finally, we consider the energy spectra of $\nu_e$ and ${\bar \nu}_e$ outflowing from the inner ejecta of a type II SN explosion \cite{janka}.
The aim is to analyze the effects of gravity (gravitational redshift) on the energy spectra \cite{fuller}.

Let us define the ${\nu}_e$ neutrinosphere at  $r^{\nu-sp}_{\nu_e}$ and the $\overline{\nu}_e$ neutrinosphere at $r^{\nu-sp}_{\overline{\nu}_e} $.
The electron fraction $Y_e$ can be written as
\[
Ye = \frac{1}{ 1 + R_{\frac{n}{p}} }, \hskip 0.7truecm R_{\frac{n}{p}} \equiv
R^0_{\frac{n}{p}} \cdot \Gamma, \quad
R^0_{\frac{n}{p}} \simeq \left[\frac{ L^{\nu-sp}_{\overline{\nu}_e } \, \langle
E^{\nu-sp}_{\overline{\nu}_e} \rangle}{ L^{\nu-sp}_{\nu_e} \, \langle
E^{\nu-sp}_{\nu_e}\rangle}
\right]
\,,\,\,\Gamma \equiv \bigg[ \frac{g_{00}(r^{\nu-sp}_{\overline{\nu}_e})}{g_{00}(r^{\nu-sp}_{{\nu}_e})} \bigg]^{3/2}\,.
\]
In these equations, $ L^{\nu-sp}_{\overline{\nu}_e},\langle E^{\nu-sp}_{\overline{\nu}_e}
\rangle$ are the averaged $\overline{\nu}_e$ energy and luminosity as measured by a
locally inertial observer at rest at the $\overline{\nu}_e$ neutrinosphere (the quantities characterizing the ${\nu}_e$ energy and luminosity at the ${\nu}_e$ neutrinosphere are similarly defined). The $Y_e$ results are shown in Fig. 2.

\begin{figure}[h!]
\centering
 \includegraphics[height=10.cm,width=18.cm]{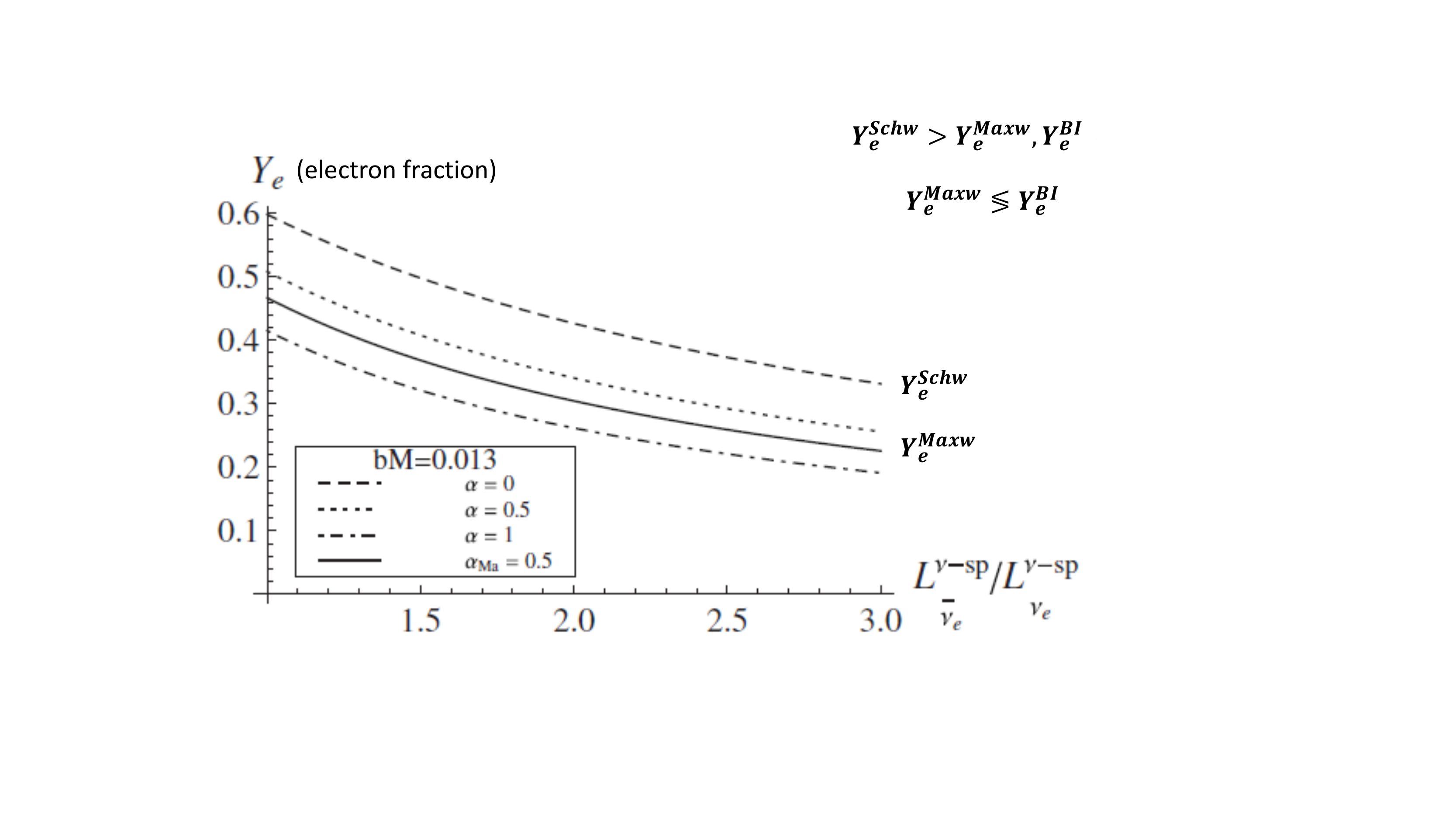}
 \label{Fig2} \vspace{-2.7cm}
  \caption{Electron fraction $Y_e$ vs the antineutrino/neutrino luminosity ratio. We compare the Schwarzschild electron fraction with the ones coming from Maxwell and Born-Infeld theories.}
 \end{figure}



\end{document}